\documentclass[reprint,amsmath,amssymb,aps,floatfix,groupedaddress,superscriptaddress,nofootinbib,preprintnumbers]{revtex4-1}
\usepackage{float}
\usepackage{amsmath}
\usepackage[pdftex]{hyperref}
\usepackage{graphicx,color}
\usepackage{dcolumn}
\usepackage{bm}
\usepackage{here}
\usepackage{comment}
\usepackage[paperwidth=210mm,paperheight=297mm,centering,hmargin=1.5cm,vmargin=2.0cm]{geometry}
\usepackage{tabularx}

\allowdisplaybreaks[1]

\makeatletter
\newsavebox{\@brx}
\newcommand{\llangle}[1][]{\savebox{\@brx}{\(\m@th{#1\langle}\)}%
  \mathopen{\copy\@brx\mkern2mu\kern-0.9\wd\@brx\usebox{\@brx}}}
\newcommand{\rrangle}[1][]{\savebox{\@brx}{\(\m@th{#1\rangle}\)}%
  \mathclose{\copy\@brx\mkern2mu\kern-0.9\wd\@brx\usebox{\@brx}}}
\makeatother

\begin{document}

\preprint{J-PARC-TH-0220}

\title{ Pileup corrections on higher-order cumulants } 

\author{Toshihiro Nonaka}
\email{nonaka.toshihiro.ge@u.tsukuba.ac.jp}
\affiliation{Tomonaga\,Center\,for\,the\,History\,of\,the\,Universe,\,University\,of\,Tsukuba,\,Tsukuba,\,Ibaraki\,305,\,Japan}
\author{Masakiyo Kitazawa}
\email{kitazawa@phys.sci.osaka-u.ac.jp}
\affiliation{Department\,of\,Physics,\,Osaka\,University,\,Toyonaka,\,Osaka 560-0043,\,Japan}
\affiliation{J-PARC Branch,\,KEK Theory Center,\,Institute\,of\,Particle\,and\,Nuclear\,Studies,\,KEK,\,203-1,\,Shirakata,\,Tokai,\,Ibaraki,\,319-1106,\,Japan}
\author{ShinIchi Esumi}
\email{esumi.shinichi.gn@u.tsukuba.ac.jp}
\affiliation{Tomonaga\,Center\,for\,the\,History\,of\,the\,Universe,\,University\,of\,Tsukuba,\,Tsukuba,\,Ibaraki\,305,\,Japan}

\begin{abstract}
We propose a method to remove the contributions of pileup events from 
higher-order cumulants and moments of event-by-event particle distributions.
Assuming that the pileup events are given by the superposition of two
independent single-collision events, we show that the 
true moments in each multiplicity bin can be 
obtained recursively from lower multiplicity events. 
In the correction procedure
the necessary information 
are only the probabilities of pileup events.
Other terms are extracted from the experimental data.
We demonstrate that the true cumulants can be reconstructed successfully
by this method in simple models.
Systematics on trigger inefficiencies and correction parameters are discussed. 
\end{abstract}
\maketitle

\newcommand{\ave}[1]{\ensuremath{\langle#1\rangle} }
\onecolumngrid 
\section{Introduction}
One of the ultimate goals of high energy physics experiments is to study 
the Quantum Chromo-Dynamics (QCD) phase diagram and especially 
the search for the QCD critical point~\cite{Bluhm:2020mpc}.
It was suggested that the higher-order fluctuation observables 
are sensitive to the critical point, and the phase transition from
quark-gluon plasma phase to the hadron-gas
phase~\cite{susceptibility,correlation,Asakawa:2009aj,Friman}.
There have been lots of experimental efforts to measure the higher-order cumulants 
of event-by-event net-particle distributions such as net-proton, net-charge and net-kaon 
multiplicity distributions reported by ALICE~\cite{Arslandok:2020mda}, 
HADES~\cite{Adamczewski-Musch:2020slf}, NA61~\cite{Mackowiak-Pawlowska:2020glz} and 
STAR collaborations~\cite{Aggarwal:2010wy,net_charge,net_proton,Adamczyk:2017wsl,Adam:2020unf,Nonaka:2020crv}.
In particular, the ratio of fourth to the second order cumulants of the net-proton distributions 
were presented to behave nonmonotonically as a function of collision energy
with a strong enhancement at $\sqrt{s_{\rm NN}}=$~7.7~GeV~\cite{Adam:2020unf}.
This result is qualitatively similar to 
a theoretical model prediction~\cite{Stephanov:2011pb}, 
which would imply the existence of the critical point 
at low collision energy region.
In order to establish the signal from the critical point, it is important to investigate further 
lower collision energy region, where the signal is predicted to decrease 
again~\cite{Stephanov:2011pb}.
Such experiments are being carried out by the STAR collaboration with the fixed-target mode 
instead of the collider mode at RHIC.
In addition, future facilities focusing on
low collision energies $\sqrt{s_{\rm NN}}<$10~GeV 
like CBM~\cite{Friman:2011zz} and J-PARC-HI~\cite{J-PARC-HI} experiments 
are also going to run with fixed-target mode.

One major issue expected in fixed-target experiments is pileup events.
When two collision events occur on the target within 
a small space and and time interval,
they are identified as a single event.
These events are called the pileup event.
Usually,
the rate of the pileup events is well suppressed and the effect 
is negligible for most of the measurements. 
Even if not, the effect would be removed from any averaged observables 
once the pileup probability is well understood and estimated.
Unfortunately, this is not the case for the higher order
fluctuation observables. 
It was pointed out that the pileup events lead to a strong enhancement of 
the fourth order cumulant and moment at central collisions~\cite{Sombun:2017bxi,Garg:2017agr}.
However, the correction method has not been known.
Because the pileup events give rise to
a fake enhancement, this effect 
makes it difficult to interpret the final results of the critical point search.
In the future experiments at CBM and J-PARC-HI, a proper understanding 
of the pileup events will be more crucial, because
high collision rates which will be achieved by these experiments
would enhance the probability of the pileups.
Development of a correction method are urgent for proper understanding 
of upcoming experimental results.

In the present work, we propose a method to correct higher-order moments and cumulants for the pileup effects. 
  Assuming that the pileup events consist of independent single-collision
  events,
  we derive the relations to connect the experimentally-observed moments
  including the pileup effects with the true moments.
  We also propose a systematic procedure to obtain the true cumulants
  using these relations by a recursive reconstruction
  of moments
  from lower multiplicity events.
  We then demonstrate the validity of this method by applying it to
  simple models.
  Systematics on trigger inefficiencies and correction parameters
  are also discussed. 

This paper is organized as follows. In Sec.~\ref{sec:method}, we explain the methodology 
for pileup corrections and derive correction formulas.
The method is demonstrated in Sec.~\ref{sec:model} with the extreme cases and realistic situations.
In Sec.~\ref{sec:pu_sensitivity} we discuss systematics of our method.
We then summarize this work in Sec.~\ref{sec:summary}.

\section{Methodology\label{sec:method}}

\subsection{Pileup events}

Let us first clarify the definition of the pileup events and assumptions
to be made in the present study.

First, in experimental analyses the pileup events are removed
using various methods, some of which are carried out by offline analysis.
For example, suppose a correlation plot between number of particles
measured by two detectors in different acceptances.
The normal single-collision events are expected to appear as a band
having a positive or negative slope. 
On the other hand, the pileup events would appear as additional bands 
having different slopes and/or offsets, while there could be
be some uncorrelated components.
The pileup events are then 
removed by cutting outlier events outside the correlated band.
However, there will be a finite probability that the band 
is contaminated by the pileup events due to randomness.
These events cannot be removed by this analysis.
We call these residual events remaining after various cutting
as the pile up events, and investigate the correction of 
their effects.

Second, in the following discussion two distribution functions play
crucial roles.
One of them is the ``multiplicity distribution'',
i.e. the distribution of the number of produced particles 
measured at mid- or forward-rapidity.
The multiplicity is sometimes used to define
centrality.
The other distribution is the ``particle distribution''.
In the event-by-event analysis, we focus on the distribution of 
the number of a specific particle or charge, $N$, such as the
net-proton number or net-charge, represented by the probability
distribution function $P(N)$, and study its cumulants.
Throughout this study, we consider the distribution of a single variable
$N$ to simplify the discussion,
but it is straightforward to extend the following method to deal with the
multi-particle distributions, $P(N_1,N_2,\cdots)$, and the mixed cumulants
of various particle species.

The distribution $P(N)$ depends on the multiplicity.
Throughout this paper, we denote the 
experimentally-observed distribution function including the pileup effects
at multiplicity $m$ as $P_m(N)$, while 
the distribution of $N$ in true single-collision events are
represented as $P_m^{\rm t}(N)$.
We suppose that $m$ and $N$ would be
measured at different 
acceptances to reduce the auto-correlation effects between $m$ and $N$.
This means that a collision event with $m=0$ can take place
and have with nonzero $N$.

Third, except for Sec.~\ref{sec:3} we consider the pileup events
composed of two single-collision events.
As discussed in Sec.~\ref{sec:3}, it is possible to extend 
the following analysis to include the pileup events with more than two
single-collisions.
The probability of those events,
however, are usually well suppressed and
explicit consideration of their effects are not needed.
The important assumption taken throughout this paper 
is that two single-collision events included in a pileup event are independent.

\subsection{Pileup correction}

Let us suppose that the pileup events occur with the probability
$\alpha_m$ at the $m$th multiplicity bin.
Then, the probability to find $N$ particles of interest at 
multiplicity $m$ with the pileup effects is given by
\begin{eqnarray}
	P_{m}(N) &=& (1-\alpha_{m})P^{\rm t}_{m}(N) + \alpha_m {P}^{\rm pu}_{m}(N), \label{eq:mosp}
\end{eqnarray}
where $P^{\rm t}_{m}(N)$ and $P^{\rm pu}_{m}(N)$ are the probability
distribution functions of $N$ for the true (single-collision) and pileup
events, respectively.
The pileup events are further decomposed into
the ``sub-pileup'' events given by the superposition of
two single-collision events with multiplicities $i$ and $j$ satisfying
$m=i+j$ as 
\begin{eqnarray}
	P^{\rm pu}_{m}(N) &=& \sum_{i,j}\delta_{m,i+j}w_{i,j}P^{\rm sub}_{i,j}(N), \label{eq:pu} \\
	P^{\rm sub}_{i,j}(N) &=& 
	\sum_{N_{i},N_{j}}\delta_{N,N_{i}+N_{j}}P^{\rm t}_{i}(N_{i})P^{\rm t}_{j}(N_{j}), \label{eq:minipu}
\end{eqnarray}
where $P^{\rm sub}_{i,j}(N)$ represents the probability distribution 
of $N$ in the sub-pileup events labeled by $(i,j)$,
and $w_{i,j}$ is the probability to observe the sub-pileup events 
among the pileup events at the $m$th multiplicity bin.
The sum over $i$ and $j$ runs non-negative integers.
Obviously $w_{i,j}$ satisfies $w_{i,j}=w_{j,i}$ and 
\begin{eqnarray}
  \sum_{i,j} \delta_{m,i+j} w_{i,j} = 1. \label{eq:w}
\end{eqnarray}
From Eq.~\ref{eq:w} one also finds $w_{0,0}=1$.

The pileup probabilities $\alpha_m$ and $w_{i,j}$ are related to
the multiplicity distribution of the single-collision events.
Let $T(m)$ be the multiplicity distribution, i.e. probability that
a collision event with multiplicity $m$ occurs for 
all single-collision events.
When all sub-pileup events are rejected by an experimental analysis
with the same resolution, 
the probability to find a sub-pileup event labeled by $(i,j)$
among all collision events is given by $\alpha T(i) T(j)$, 
where $\alpha$ denotes the probability to find a pileup event among
all collision events.
Also, the probability to find an event with multiplicity $m$
without distinction between single-collision and pileup events
is given by $(1-\alpha) T(m) + \alpha \sum_{i,j}\delta_{m,i+j}T(i)T(j)$.
We thus have
\begin{eqnarray}
  w_{i,j} &=& \frac{\alpha T(i)T(j)}{\sum_{i,j}\delta_{m,i+j}\alpha T(i)T(j)},
  \label{eq:w_ij}
  \\
  \alpha_{m} &=& \frac{ \alpha \sum_{i,j}\delta_{m,i+j}T(i)T(j) }
        { (1-\alpha) T(m) + \alpha \sum_{i,j}\delta_{m,i+j}T(i)T(j) }.
  \label{eq:alpha_m}
\end{eqnarray}
Therefore, in this case $\alpha_m$ and $w_{i,j}$ are completely determined
from $\alpha$ and $T(m)$.
We, however, note that Eqs.~\ref{eq:w_ij} and \ref{eq:alpha_m} might not
hold in realistic experimental cases 
if the probability distribution of the pileup rejection is different from the multiplicity distribution. 
  We thus do not use Eqs.~\ref{eq:w_ij} and \ref{eq:alpha_m} explicitly
  in the rest of this section.
  In real experiments, $w_{i,j}$ can directly be estimated by some
  reasonable assumptions within the experimental simulation.
  The models employed in Sec.~\ref{sec:model} satisfy
  Eqs.~\ref{eq:w_ij} and \ref{eq:alpha_m}.

From Eqs.~\ref{eq:mosp}, \ref{eq:pu} and \ref{eq:minipu}, 
the moment generating function~\cite{Asakawa:2015ybt} for events
at
multiplicity $m$ is expressed as
\begin{eqnarray}
  G_{m}(\theta) &=& \sum_{N}e^{N\theta}P_{m}(N)
  \nonumber \\
  &=& ( 1 - \alpha_m ) G^{\rm t}_{m}(\theta) + 
  \alpha_m \sum_{i,j}\delta_{m,i+j} w_{i,j} G^{\rm sub}_{i,j}(\theta),
\end{eqnarray}
with
\begin{eqnarray}
        G^{\rm sub}_{i,j}(\theta) = G^{\rm t}_{i}(\theta)G^{\rm t}_{j}(\theta),
\end{eqnarray}
where $G^{\rm t}_{m}(\theta)=\sum_N e^{N\theta} P^{\rm t}_m(N)$ is the moment generating
function of $P_m^{\rm t}(N)$.
The $r$th order moment of the observed distribution $P_{m}(N)$ is then given by
\begin{eqnarray}
  \langle N^{r}\rangle_{m}
  &=& \sum_N N^r P_m(N)
  = \frac{d^{r}}{d\theta^{r}}G(\theta)|_{\theta=0} \nonumber \\
	&=& (1-\alpha_{m})\langle N^{r}\rangle_{m}^{\rm t}
        + \alpha_m \sum_{i,j}\delta_{m,i+j}w_{i,j}\ave{N^{r}}_{i,j}^{\rm sub},
        \label{eq:prob}
\end{eqnarray}
with $\langle N^{r}\rangle_{m}^{\rm t}= \sum_N N^r P_m^{\rm t}(N)$ and 
\begin{eqnarray}
  \ave{N^{r}}_{i,j}^{\rm sub} &=&
    \sum_N N^r P_{i,j}^{\rm sub}(N) =
  \sum^{r}_{k=0}\binom{r}{k}\ave{N^{r-k}}_{i}^{\rm t}\ave{N^{k}}_{j}^{\rm t}. \label{eq:sub}
\end{eqnarray}
The right-hand sides in Eq.~\ref{eq:sub} up to fourth order are written as
\begin{eqnarray}
	\ave{N}_{i,j}^{\rm sub} &=& \ave{N}_{i}^{\rm t} + \ave{N}_{j}^{\rm t}, \label{eq:muadd1} \\
	\ave{N^{2}}_{i,j}^{\rm sub} &=& \ave{N^{2}}_{i}^{\rm t} + \ave{N^{2}}_{j}^{\rm t} 
	+ 2\ave{N}_{i}^{\rm t}\ave{N}_{j}^{\rm t}, \label{eq:muadd2}\\
	\ave{N^{3}}_{i,j}^{\rm sub} &=& \ave{N^{3}}_{i}^{\rm t} + \ave{N^{3}}_{j}^{\rm t} 
	+ 3\ave{N^{2}}_{i}^{\rm t}\ave{N}_{j}^{\rm t} + 3\ave{N}_{i}^{\rm t}\ave{N^{2}}_{j}^{\rm t}, \label{eq:muadd3}\\
	\ave{N^{4}}_{i,j}^{\rm sub} &=& \ave{N^{4}}_{i}^{\rm t} + \ave{N^{4}}_{j}^{\rm t} 
	+ 4\ave{N^{3}}_{i}^{\rm t}\ave{N}_{j}^{\rm t} + 4\ave{N}_{i}^{\rm t}\ave{N^{3}}_{j}^{\rm t} 
	+ 6\ave{N^{2}}_{i}^{\rm t}\ave{N^{2}}_{j}^{\rm t}. \label{eq:muadd4}
\end{eqnarray}
We note that Eq.~\ref{eq:sub} is alternatively expressed using cumulants
in a compact form as~\cite{Asakawa:2015ybt}
\begin{equation}
  \ave{N^{r}}^{\rm sub}_{i,j, \rm c}
  = \ave{N^{r}}^{\rm t}_{i, \rm c} + \ave{N^{r}}^{\rm t}_{j, \rm c},
\end{equation}
where $\ave{N^{r}}^{\rm sub}_{i,j, \rm c}$
and $\ave{N^{r}}^{\rm t}_{j, \rm c}$ are the cumulants of the sup-pileup
and true distributions, respectively.

Substituting Eq.~\ref{eq:sub} into Eq.~\ref{eq:prob},
one obtains formulas connecting 
$\langle N^{r}\rangle_{m}$ and $\langle N^{r}\rangle_{m}^{\rm t}$.
It is notable that in these formulas the observed moment
$\langle N^{r}\rangle_{m}$ is given
by the combination of the true moments $\langle N^{r'}\rangle_{m'}^{\rm t}$
with $r'\le r$ and $m'\le m$.

The true moments $\langle N^{r}\rangle_{m}^{\rm t}$ are obtained
from the observed moments $\langle N^{r}\rangle_{m}$
by solving Eqs.~\ref{eq:prob} and \ref{eq:sub}.
This procedure can be carried out recursively starting from
$m=0$ and $r=1$, and by increasing $m$ and $r$.
To see this, it is convenient to rewrite Eqs.~\ref{eq:prob} and
\ref{eq:sub} as 
\begin{eqnarray}
	\ave{N^{r}}_{m}^{\rm t} &=& \cfrac{\ave{N^{r}}_{m}-\alpha_{m}C_{m}^{(r)}}{1-\alpha_{m}+2\alpha_m w_{m,0}}, \label{eq:final}
\end{eqnarray}
with 
\begin{eqnarray}
	C_{m}^{(r)} &=& \mu_{m}^{(r)} + \sum_{i,j>0}\delta_{m,i+j}w_{i,j}\ave{N^{r}}_{i,j}^{\rm sub}, \label{eq:final_corr}
\end{eqnarray}
and 
\begin{eqnarray}
  \mu_m^{(r)} =
  \begin{cases}
    \displaystyle
    2w_{m,0}\sum^{r-1}_{k=0}\binom{r}{k}\ave{N^{r-k}}_{0}^{\rm t}\ave{N^{k}}_{m}^{\rm t}
    & (m>0), \\
    \displaystyle
    \sum^{r-1}_{k=1}\binom{r}{k}\ave{N^{r-k}}_{0}^{\rm t}\ave{N^{k}}_{0}^{\rm t}
    & (m=0).
  \end{cases}
  \label{eq:mu_m}
\end{eqnarray}
Up to the fourth order, the explicit forms of $\mu_{m}^{(r)}$ are 
\begin{eqnarray}
	\mu_{m}^{(1)} &=& 0, \label{eq:mu_nonzerom1}\\
	\mu_{m}^{(2)} &=& 2w_{m,0}\left[\ave{N^{2}}_{0}^{\rm t}+2\ave{N}_{m}^{\rm t}\ave{N}_{0}^{\rm t}\right], \label{eq:mu_nonzerom2}\\
	\mu_{m}^{(3)} &=& 2w_{m,0}\left[\ave{N^{3}}_{0}^{\rm t}+3\ave{N^{2}}_{m}^{\rm t}\ave{N}_{0}^{\rm t}
	+3\ave{N}_{m}^{\rm t}\ave{N^{2}}_{0}^{\rm t}\right], \label{eq:mu_nonzerom3}\\
	\mu_{m}^{(4)} &=& 2w_{m,0}\left[\ave{N^{4}}_{0}^{\rm t}+4\ave{N^{3}}_{m}^{\rm t}\ave{N}_{0}^{\rm t}
	+4\ave{N}_{m}^{\rm t}\ave{N^{3}}_{0}^{\rm t}+6\ave{N^{2}}_{m}^{\rm t}
	\ave{N^{2}}_{0}^{\rm t}\right],\label{eq:mu_nonzerom4}
\end{eqnarray}
for $m>0$ and
\begin{eqnarray}
	\mu_{0}^{(1)} &=& 0, \label{eq:mu_zerom1}\\
	\mu_{0}^{(2)} &=& 2\ave{N_{0}}^{\rm t}\ave{N_{0}}^{\rm t}, \label{eq:mu_zerom2}\\
	\mu_{0}^{(3)} &=& 6\ave{N_{0}^{2}}^{\rm t}\ave{N_{0}}^{\rm t},\label{eq:mu_zerom3} \\
	\mu_{0}^{(4)} &=& 8\ave{N_{0}^{3}}^{\rm t}\ave{N_{0}}^{\rm t}+6\ave{N_{0}^{2}}^{\rm t}
	\ave{N_{0}^{2}}^{\rm t}. \label{eq:mu_zerom4}
\end{eqnarray}

To obtain the true moments $\ave{N^r}_m^{\rm t}$, 
we first use the fact that $C_0^{(1)}=0$, which leads to 
$\ave{N}_0^{\rm t}=\ave{N}_0/(1+\alpha_0)$.
Next, Eqs.~\ref{eq:final_corr} and \ref{eq:mu_m}
shows that the correction factors $C_0^{(r)}$ at $m=0$ are 
given only by the moments $\ave{N^{r'}}_{0}^{\rm t}$ with
$r'<r$.
One thus can obtain $\ave{N^r}_0^{\rm t}$ recursively 
from lower order
up to any higher orders.
Similarly, one can obtain the true moments at multiplicity $m=1$
from lower order moments up to any order using the fact that
the correction factor $C_1^{(r)}$ consists of 
$\ave{N^{r'}}_{m'}^{\rm t}$ with $r'\le r$ and $m'\le m$.
By repeating the same procedure 
one can obtain the true moments for all multiplicities.

An important remark here is that this procedure can be carried out
in almost data-driven way. 
Only thing we need is the probabilities $w_{i,j}$ and $\alpha_m$,
which would be determined by simulations.

\subsection{Pileups composed of more than two single-collision events}
\label{sec:3}

So far, we considered the pileup events
composed of two single-collision events.
It is not difficult to extend these results to include the 
pileup events composed of three single-collision events.
In this case, Eq.~\ref{eq:pu} is modified 
as 
\begin{eqnarray}
  P_m^{\rm pu} = \sum_{i,j} \delta_{m,i+j} w_{i,j} P_{i,j}^{\rm sub}(N)
  + \sum_{i,j,k} \delta_{m,i+j+k} w_{i,j,k} P_{i,j,k}^{\rm sub}(N),
  \label{eq:pu3}
\end{eqnarray}
where $P_{i,j,k}^{\rm sub}(N)$ represents the probability distribution of $N$
on the sub-pileup events composed of three single collisions with
multiplicities $i$, $j$, and $k$, and $w_{i,j,k}$ is the probability
of the sub-pileup event.
From the independence of the individual collisions,
$P_{i,j,k}^{\rm sub}(N)$ is given by
\begin{eqnarray}
  P_{i,j,k}^{\rm sub} = \sum_{N_i,N_j,N_k} \delta_{N,N_i+N_j+N_k}
  P_i^{\rm t}(N_i) P_j^{\rm t}(N_j) P_j^{\rm t}(N_j).
  \label{eq:subpu3}
\end{eqnarray}
Then, it is straightforward to derive the relations like
Eqs.~\ref{eq:prob} and \ref{eq:final}.
These results allow us to obtain the true moments $\ave{N^r}_m^{\rm t}$
recursively from small $m$ as before.
In this way, pileups with arbitrary many single-collision events
can be taken into account in principle.

\section{Model\label{sec:model}}

  In this section we apply the procedure introduced in the previous section
  to the pileup correction in simple models and demonstrate that
  the true cumulants are successfully obtained.

\subsection{Multiplicity distributions}
\label{sec:multiplicity}

Let us first generate a realistic multiplicity distribution with
pileup events.
We employ the Glauber and two-component model for this purpose.
Two gold nuclei are collided in the Glauber model, where the $pp$ cross 
section is chosen to be $33$~mb.
The number of participant nucleons, $N_{\rm part}$, and binary collisions, $N_{\rm coll}$, 
are obtained.
In order to propagate $N_{\rm part}$ and $N_{\rm coll}$ to the multiplicity, 
we define the number of sources,
$N_{\rm sc}$ as
\begin{eqnarray}
	N_{\rm sc} = (1-x)N_{\rm part} + xN_{\rm coll},
\end{eqnarray}
where $x$ is the fraction of the hard component. We
choose
$x=0.1$ for the simulation.
Particles are then generated from each source $N_{\rm sc}$ based on the negative binomial distribution:
\begin{eqnarray}
	P_{\mu,k}(N) = \cfrac{\Gamma(N+k)}{\Gamma(N+1)\Gamma(k)} \cdot \cfrac{(\mu/k)^N}{(\mu/k+1)^{N+k}},
\end{eqnarray}
where $\mu$ is the mean value of particles generated from one source, and $k$ corresponds 
to the inverse of width of the distribution.
$\mu=1.0$ and $k=1.0$ are chosen for the simulation.
In order to simulate the pileup events as well as normal single-collision events, 
multiplicities from two collision events are randomly superimposed 
with the probability $\alpha=0.05$. 
In this way, 10 million Au+Au collision events are processed.
We note that in this model the pileup probabilities $w_{i,j}$ and
  $\alpha_m$
  are given by Eqs.~\ref{eq:w_ij} and \ref{eq:alpha_m} by construction.

\begin{figure*}[htbp]
	\begin{center}
	\includegraphics[width=130mm]{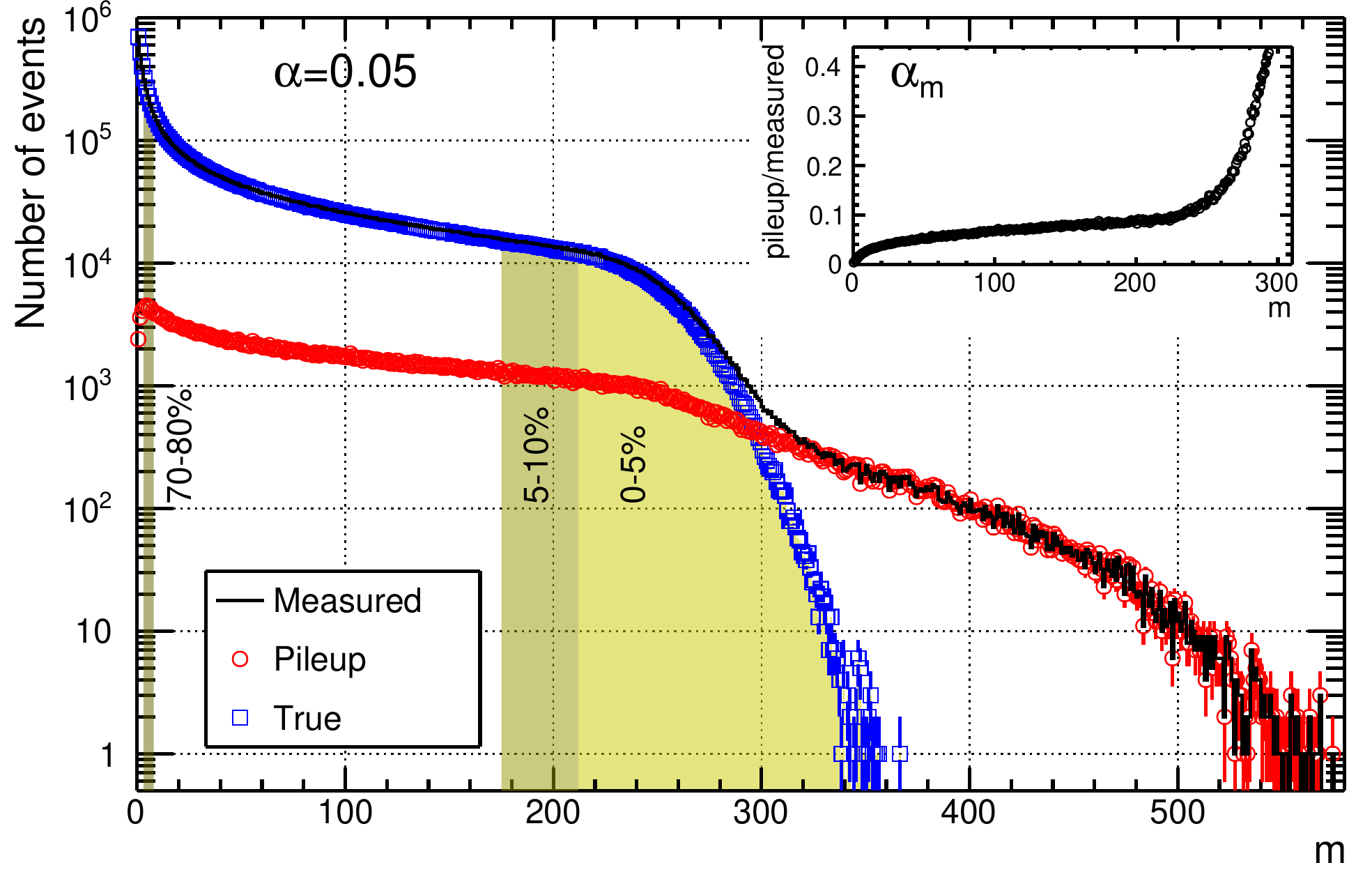}
	\end{center}
	\caption{
		The multiplicity distribution generated from the Glauber and two component model. 
		The black line includes the contribution from pileup events with $\alpha=0.05$ (measured distribution). 
		The red open circles are the distribution from pileup events, and the blue squares are from 
		the normal single collision events. 
		The bands indicate 0-5\%, 5-10\% and 70-80\% centralities.
		The inset panel shows the ratio of pileup to measured distributions 
		as a function of multiplicity ($\alpha_{m}$).
		}
	\label{fig:mult}
\end{figure*}

The resulting multiplicity distribution is shown by the black line in Fig.~\ref{fig:mult}. 
The blue squares show the multiplicity distribution from single-collision events, 
while those from pileup events are shown by the red circles.
It is found that, due to the pileup events, the measured distribution has the tail on top of 
the distribution from the single-collision events.
The inset panel shows $\alpha_m$, i.e. the ratio of the pileup events
at multiplicity $m$.
From the panel one finds that 
$\alpha_{m}$ grows with increasing $m$.
This behavior suggests that the effect of pileup events are more problematic 
in central collisions rather than peripheral collisions.

\begin{figure}[htbp]
	\begin{center}
	\includegraphics[width=110mm]{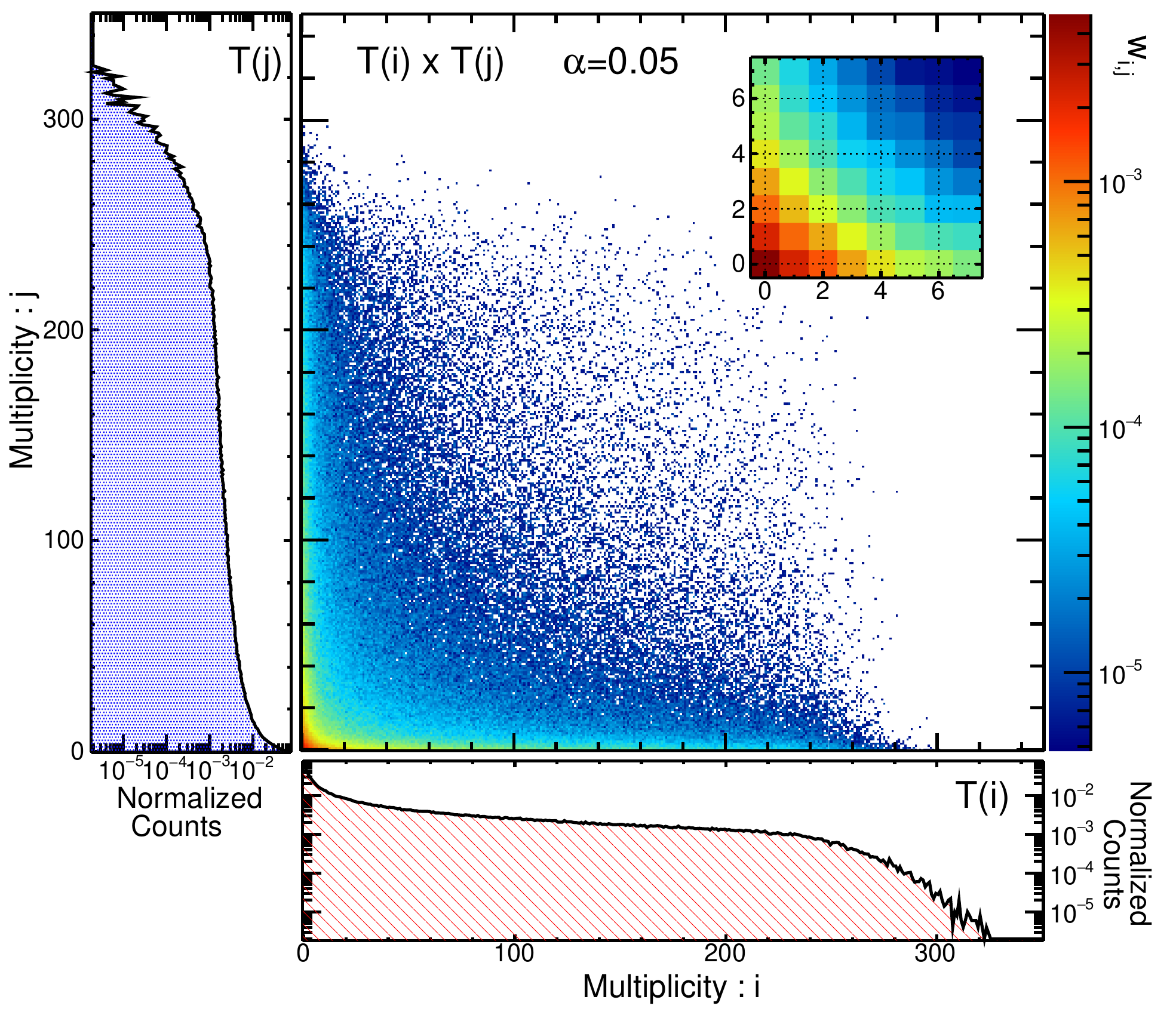}
	\end{center}
	\caption{
		Correlation between multiplicities $i$ and $j$ from two collisions 
		which forms pileup events. Z-axis is normalized by total number of events. 
		The inset panel shows the expanded plot at $x<7$ and $y<7$. 
		Projected
                histograms into x and y axes 
                are shown as well.
		}
	\label{fig:RM}
\end{figure}

  In Fig.~\ref{fig:RM}, we plot the multiplicity distribution of
  single-collision events $T(m)$ and 
  the number of sub-pileup events $(i,j)$ normalized by total simulated
  events, $\alpha T(i)T(j)$.
  From these results $w_{i,j}$ and $\alpha_m$ are constructed
  according to Eqs.~\ref{eq:w_ij} and \ref{eq:alpha_m}.
  These parameters are used in the following two subsections.


\subsection{Simple case}
\label{sec:model1}

In this and next subsections, we discuss the pileup correction
for two model distributions $P_m^{\rm t}(N)$ with the multiplicity
distribution obtained in Sec.~\ref{sec:multiplicity}.
In this subsection, we consider a simple model
where the particle number $N$ obeys the Poisson distribution 
with the mean value of $10$ at all the multiplicity bin. 
We emphasize that this model is totally impractical, 
because $10$ particles on average are created 
at both $m=0$ and $m=300$.
However, this model is suitable to demonstrate the validity of
the recursive correction procedures.
The more realistic model will be discussed in the next subsection.

\begin{figure}[htbp]
	\begin{center}
	\includegraphics[width=100mm]{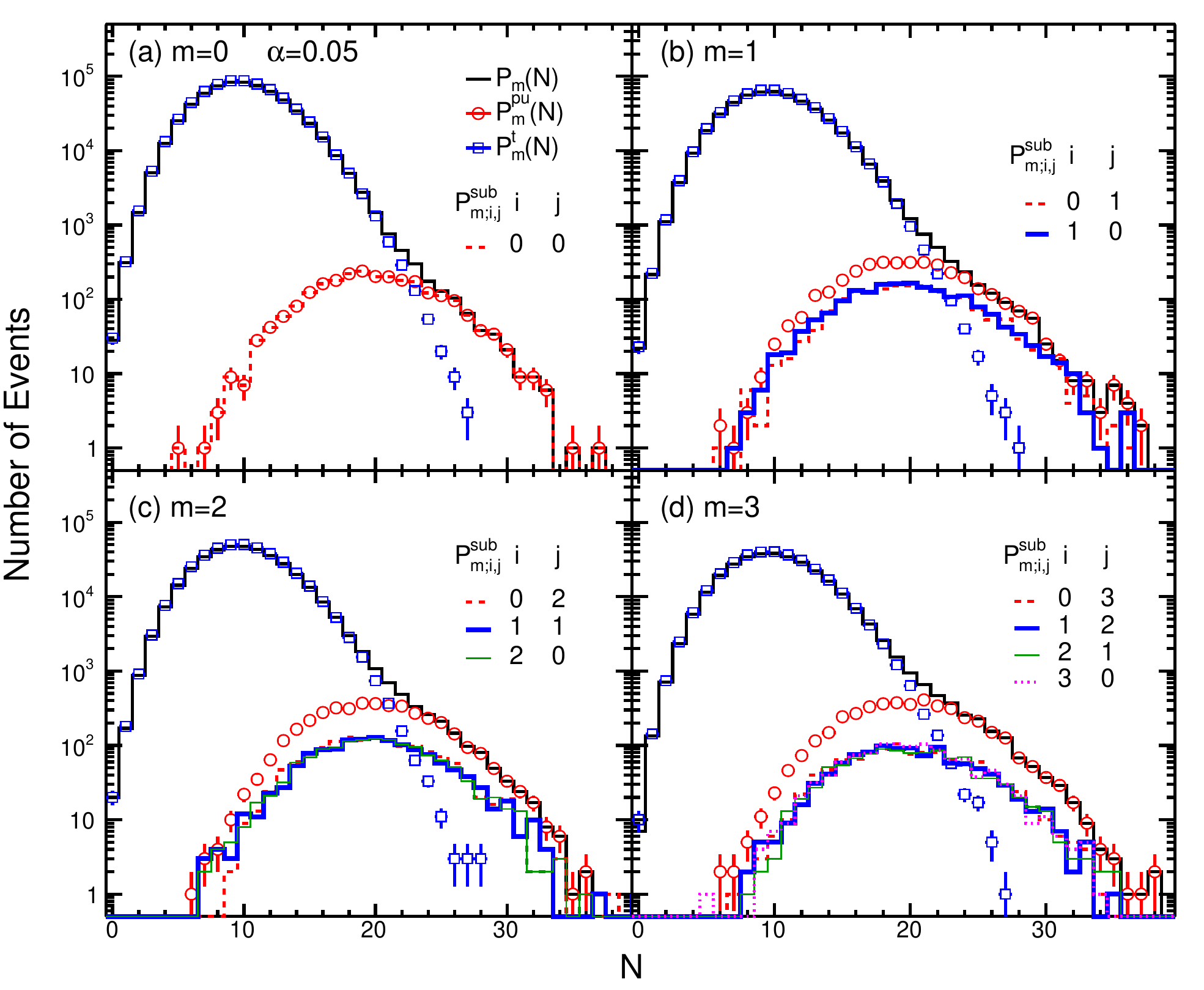}
	\end{center}
	\caption{
		Particle number distributions, $P_{m;i,j}(N)$, in the simple model in Sec.~\ref{sec:model1}
		for the first 4 multiplicity bins, (a) $m=0$, (b) $m=1$, (c) $m=2$ and (d) $m=3$. 
		The red circles shows the pileup events, 
		and blue squares are for single-collision events.
		The black solid line is for measured events.
		Distributions for sub-pileups are shown in colored or dotted lines. 
		}
	\label{fig:minipu_poisson}
\end{figure}

Figure~\ref{fig:minipu_poisson} shows the particle distribution 
for the first 4 multiplicity bins ($m=0$, 1, 2, 3).
The red circles show pileup events, and the blue squares show the single-collision events. 
The measured distribution given by the sum of these distributions is shown by
the
black solid lines, which
are
found to have 
a bump structure at
$N\gtrsim20$ due to pileup events. 
Other colored lines show the sub-pileup events for all possible combinations
of ($i$,$j$) with $m=i+j$. 
In the case of $m=0$, there is only one combination for sub-pileup,
$(i,j)=(0,0)$, and
the distributions of pileup and sub-pileup events are identical 
and $w_{0,0}=1$.
As shown in Fig.~\ref{fig:minipu_poisson} (b), (c) and (d), 
in the case of $m\ge1$, the pileup distribution consists of multiple
sub-pileup events with $(i,j)=(m,0)$, $\cdots$, $(0,m)$.

\begin{figure*}[htbp]
	\begin{center}
	\includegraphics[width=130mm]{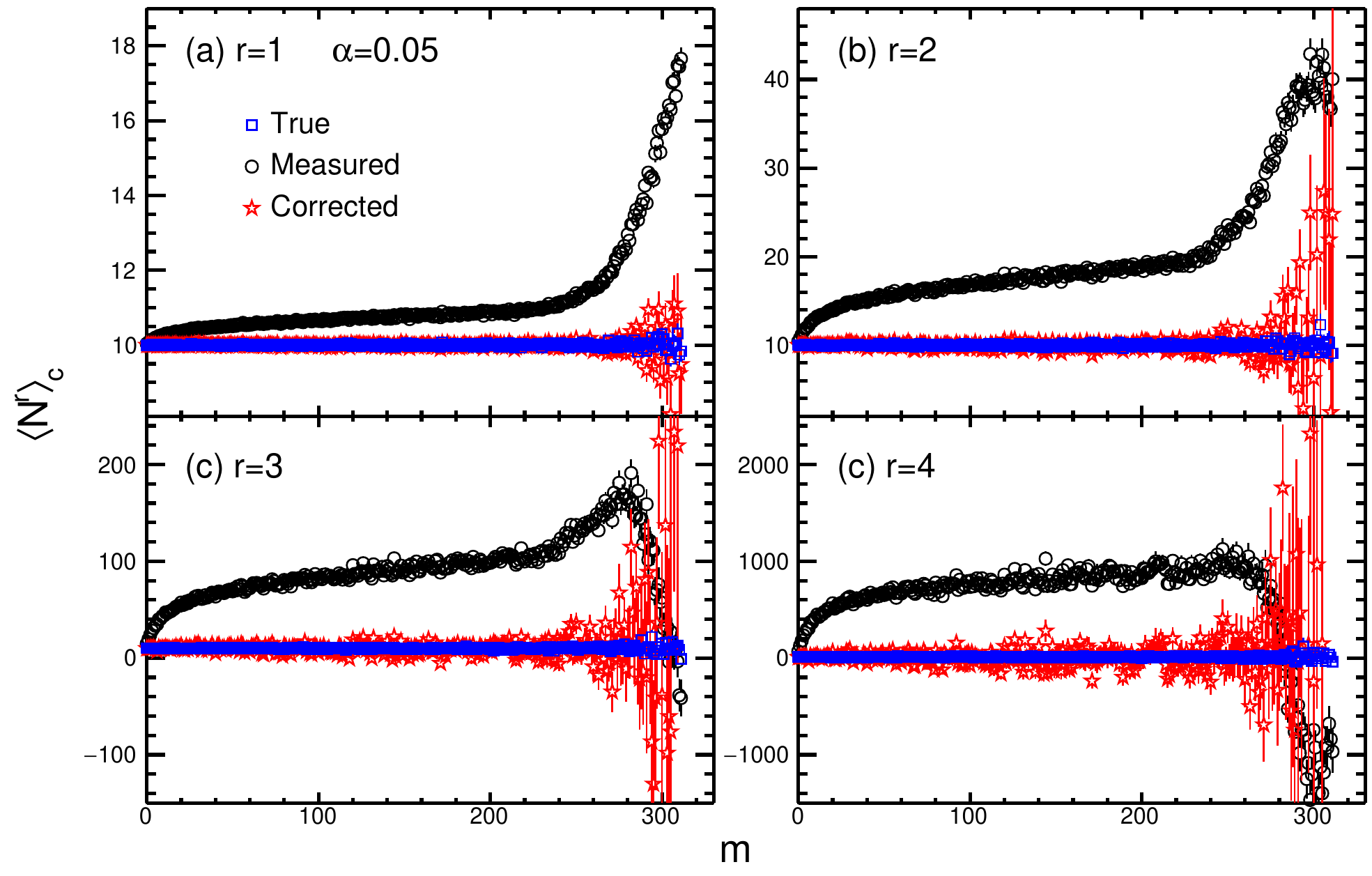}
	\end{center}
	\caption{
	  Cumulants up to the $4$th-order as a function of multiplicity
          in the simple model in Sec.~\ref{sec:model1}. 
	  True values of cumulants are shown
          by the blue squares, and the measured value of cumulants 
	  (including pileup events) are shown
          by the black circles.
	  The red stars show the results corrected for pileups.
	}
	\label{fig:extreme}
\end{figure*}

In Fig.~\ref{fig:extreme}, we show the cumulants $\ave{N^r}_{m, \rm c}$
at the $m$th multiplicity bin.
In the figure, the cumulants are plotted for the true 
single-collision
distribution obtained by the simulation, measured distribution with 
the pileup effects, and the corrected results. 
Statistical uncertainties are estimated by bootstrap.
True cumulants are $\ave{N^{r}_{m}}=10$ by definition.
The measured cumulants
have strong deviations from this value due to the pileup events~\cite{Sombun:2017bxi,Garg:2017agr}.
It is notable that the measured cumulants especially for $r=1$ and 2 behave similar to 
what we have already seen in $\alpha_{m}$ in the inset panel in Fig.~\ref{fig:mult}.
Because the particles are generated according to the Poisson distributions having the same 
mean value, the effects from the pileup events only depend on the pileup probability. 
Corrected cumulants are found to be 
consistent with the true value $\ave{N^r}_{m, \rm c}=10$
within statistics,
which indicates that our method does work well. 
Large point-by-point variations are due to the increased statistical uncertainties after the corrections. 


\subsection{Realistic case\label{sec:realistic}}
\begin{figure}[htbp]
	\begin{center}
	\includegraphics[width=100mm]{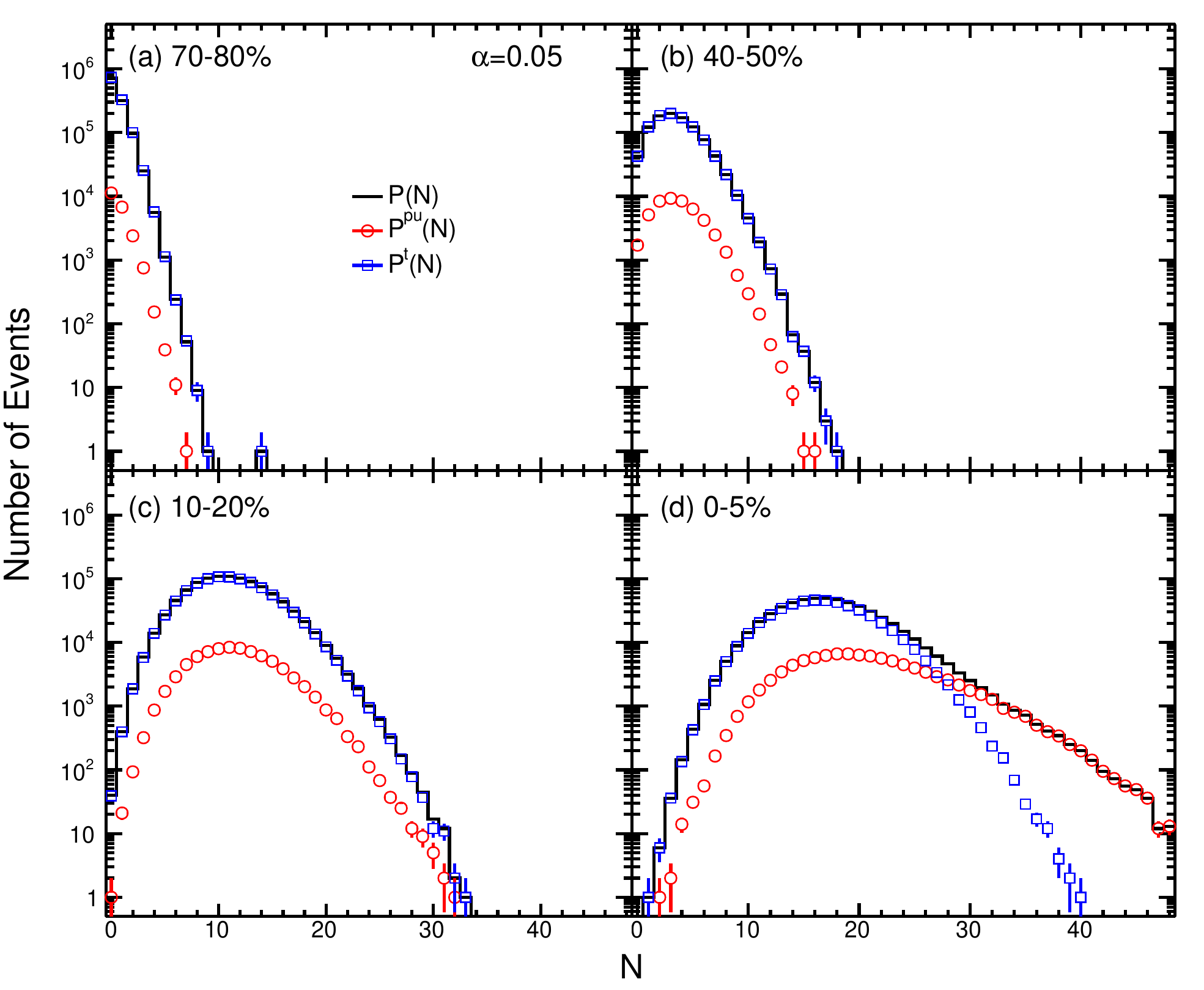}
	\end{center}
	\caption{
		Particle number distributions for 
		(a) 70-80\%, (b) 40-50\%, (c) 10-20\% and (d) 0-5\% centralities
                in the realistic model in Sec.~\ref{sec:realistic}.
		The red circles shows the pileup events, 
		and
                the
                blue squares are for single-collision events.
		The black solid line is for measured events.
		}
	\label{fig:pucent}
\end{figure}

Next, we move on to more realistic case, where the mean value of $N$ increases 
with increasing $N_{\rm part}$.
We again employ the Poisson distribution for $N$,
but in contrast to the previous subsection we assume that 
the mean value varies depending on $N_{\rm part}$
as $\ave{N}=0.05N_{\rm part}$.
We employ the same pileup probability $\alpha=0.05$.
The centrality is defined by dividing the multiplicity distribution of single-collision events 
(see Fig.~\ref{fig:mult} for corresponding regions in multiplicity distributions).
Figure~\ref{fig:pucent} shows the particle number distributions for 0-5\%, 
10-20\%, 40-50\% and 70-80\% centralities. 
The pileup distributions are found inside the true distribution at peripheral collisions, 
while the pileup distributions in central collisions appear 
as a long tail in the measured distributions. 
Thus, large effects on cumulants are expected in central collisions in this simulation~\cite{Garg:2017agr}. 

\begin{figure}[htbp]
	\begin{center}
	\includegraphics[width=130mm]{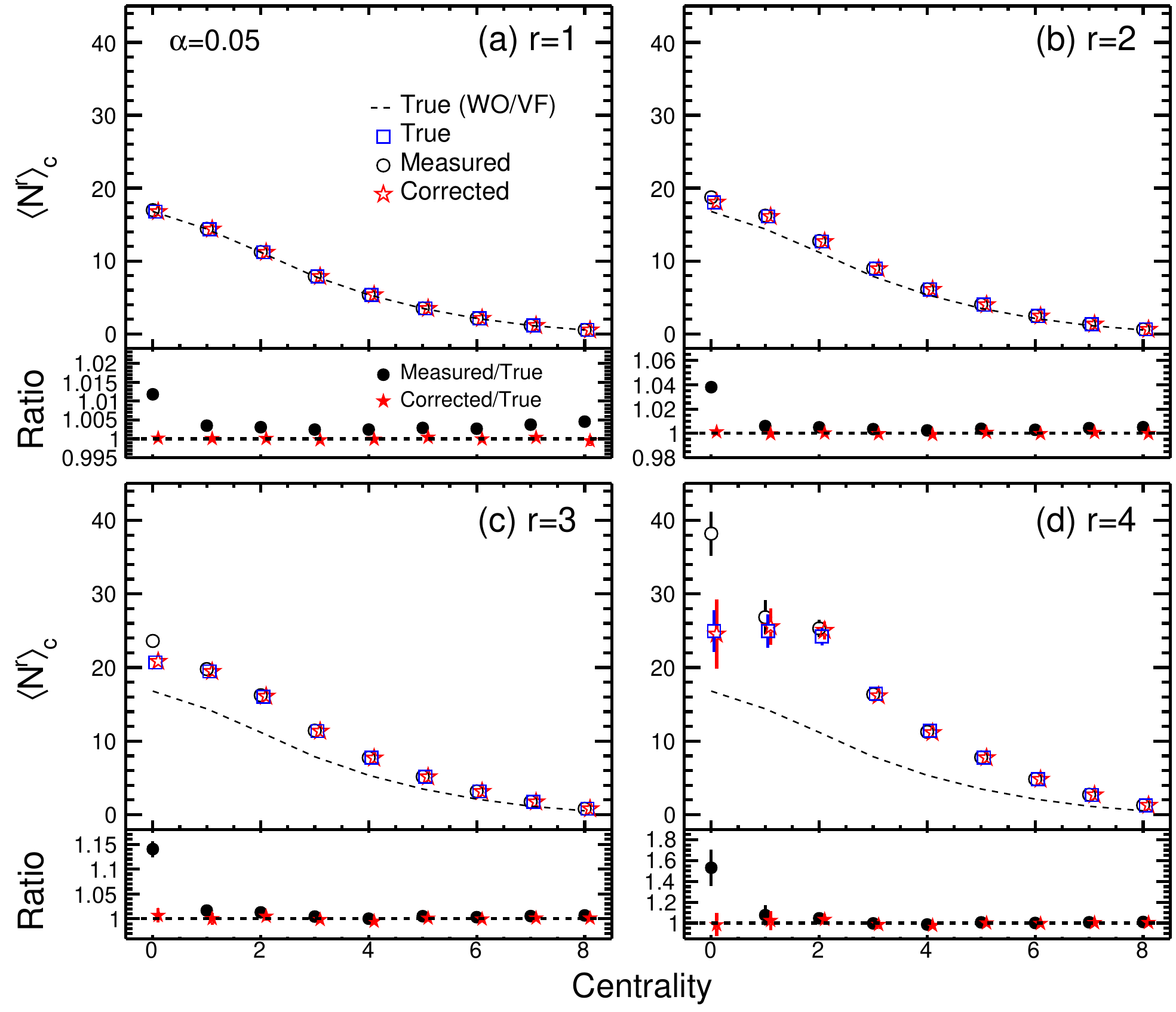}
	\end{center}
	\caption{
		Cumulants up to the $4$th-order as a function of centrality, 
		0-5\%, 5-10\%, 10-20\%, $\cdots$,
                and 70-80\% centralities from
                $x=0$ to $8$.
		True values of cumulants are shown
                by the open blue squares, 
		and the measured values of cumulants 
		(including pileup events) are shown
                by open black circles.
		The red open stars show results corrected for pileups.
		The ratios of measured and corrected results to the true 
		cumulants are shown by the filled markers in lower panels.
		The true values excluding volume fluctuations are shown in dotted lines. 
		}
	\label{fig:realistic_pu5}
\end{figure}

Cumulants for each multiplicity bin are averaged in
each centrality 
by using event statistics as a weight~\cite{Luo:2017faz}, 
which are shown in 
Fig.~\ref{fig:realistic_pu5} as a function of centrality.
The centrality is $0-5\%$, $5-10\%$, $10-20\%$... ,$70-80\%$ from $x=0$ to $x=8$.
In this case, significant deviation on measured cumulants are observed only in the central collisions, 
as was expected from Fig.~\ref{fig:pucent}.
Corrected cumulants are consistent with true cumulants
even at $0-5\%$. 
  This result shows that the pileup correction proposed
  in Sec.~\ref{sec:method} is successfully applicable to realistic
  particle distributions.

It would be interesting to discuss briefly about volume fluctuations~\cite{Skokov:2012ds}.
In Fig.~\ref{fig:realistic_pu5} 
the cumulants with fixed $N_{\rm part}$ are shown by the dotted lines.
The difference between markers and lines seen especially for
higher-order cumulants indicate the residual volume (participant) fluctuations 
even after the centrality bin width averaging~\cite{Luo:2017faz,Braun-Munzinger:2016yjz}. 
This happens because we let $N_{\rm part}$ fluctuate event by event based on the Glauber model 
and the mean value of Poisson distribution is defined as the function of $N_{\rm part}$. 
It should be noted that the location of the kink structure at $x=1\sim2$ 
(5-10\% and 10-20\% centralities), where the cumulants of $N_{\rm part}$ distribution 
have minimum or maximum values~\cite{Braun-Munzinger:2016yjz}, observed in true and corrected 
cumulants would depend on the model and the binning of the centrality. 
Interestingly, the measured cumulants including pileup events look rather 
qualitatively normal (linear) compared to the true and corrected cumulants.
This would imply that the pileup events could 
accidentally hide the characteristic kink structure 
arising from the volume fluctuations. 
One should always be careful if the effects from the volume fluctuations are removed from the measurements. 
Otherwise, the final results could be spoiled by sizable effects of both pileup and volume fluctuations. 

\section{Systematics\label{sec:pu_sensitivity}}
\subsection{Trigger inefficiency}
An important procedure of the pileup correction 
is the recursive solving of moments 
from the lowest multiplicity event at $m=0$. 
At such super-peripheral collisions, however, the event itself cannot be triggered
due to small multiplicity and the detector threshold to reject backgrounds.
The event efficiency is thus reduced in peripheral collisions, 
which is known as ``trigger inefficiency''.
  It is possible that these effects at smaller multiplicity events
  accumulate in the recursive procedure and 
  give rise to a large systematic deviation on the
  reconstructed cumulants at large $m$.

  To check this problem,
  in this subsection 
  the events for $m<20$ are artificially reduced 
  by the arbitrary function of the multiplicity, 
  and the pileup correction is not applied for this region. 
  In other words, we regard the observed moments $\ave{N^r}_m$ as
  the true moments $\ave{N^r}_m^{\rm t}$ for $m<20$, and perform the correction
  only for $m\ge20$.
  The model in Sec.~\ref{sec:realistic} is employed.
  
\begin{figure}[htbp]
	\begin{center}
	\includegraphics[width=110mm]{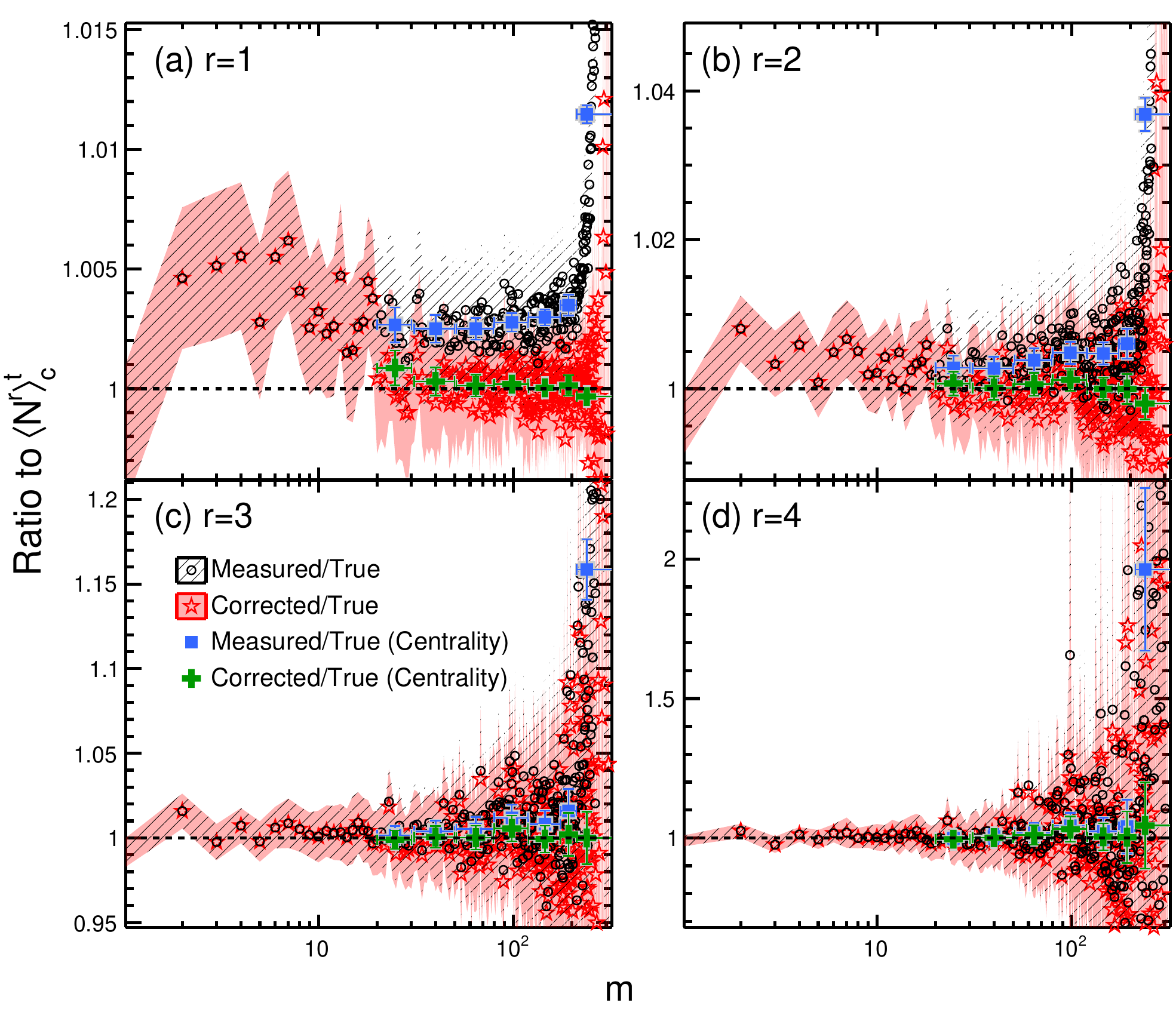}
	\end{center}
	\caption{
		The ratios of measured (black circles) and corrected (red stars) 
		cumulants with respect to the true cumulants 
		as
                functions
                of multiplicity.
		The bands represent the statistical uncertainties.
		The results averaged into 0-5, 5-10, 10-20,... and 50-60\% centralities 
		are shown in blue squares and green crosses. 
		}
	\label{fig:teff}
\end{figure}

Figure~\ref{fig:teff} shows the ratios of measured and corrected cumulants 
to the true cumulants as a function of multiplicity.
The averaged results for centrality bins
0-5, 5-10, 10-20, ... 50-60\% are also shown. 
As the correction is not applied for $m<20$, the corrected results are identical 
with measured values. On the other hand, 
the corrected cumulants for $m\ge20$ are 
quickly approaching to 
the true value,
  which shows that the correction works well regardless of incorrect correction 
factors in peripheral collisions.
This is because the sub-pileup moments (the second term in Eq.~\ref{eq:prob}) 
have less contributions from peripheral collisions due to the tiny production rate of particles of interest.
Since it depends on how significant the production of particles of interest 
is in peripheral collisions compared to central collisions, 
we would propose to check the results by changing the starting point of the recursive corrections, 
and implement it as a part of systematic uncertainties in final results.

\subsection{Correction parameters}
The new method relies on the probabilities $w_{i,j}$ and $\alpha_m$,
and other terms are all extracted from data. 
Hence, the systematic uncertainties would come from how precisely 
those parameters are determined in the simulations. 

\begin{figure}[htbp]
	\begin{center}
	\includegraphics[width=130mm]{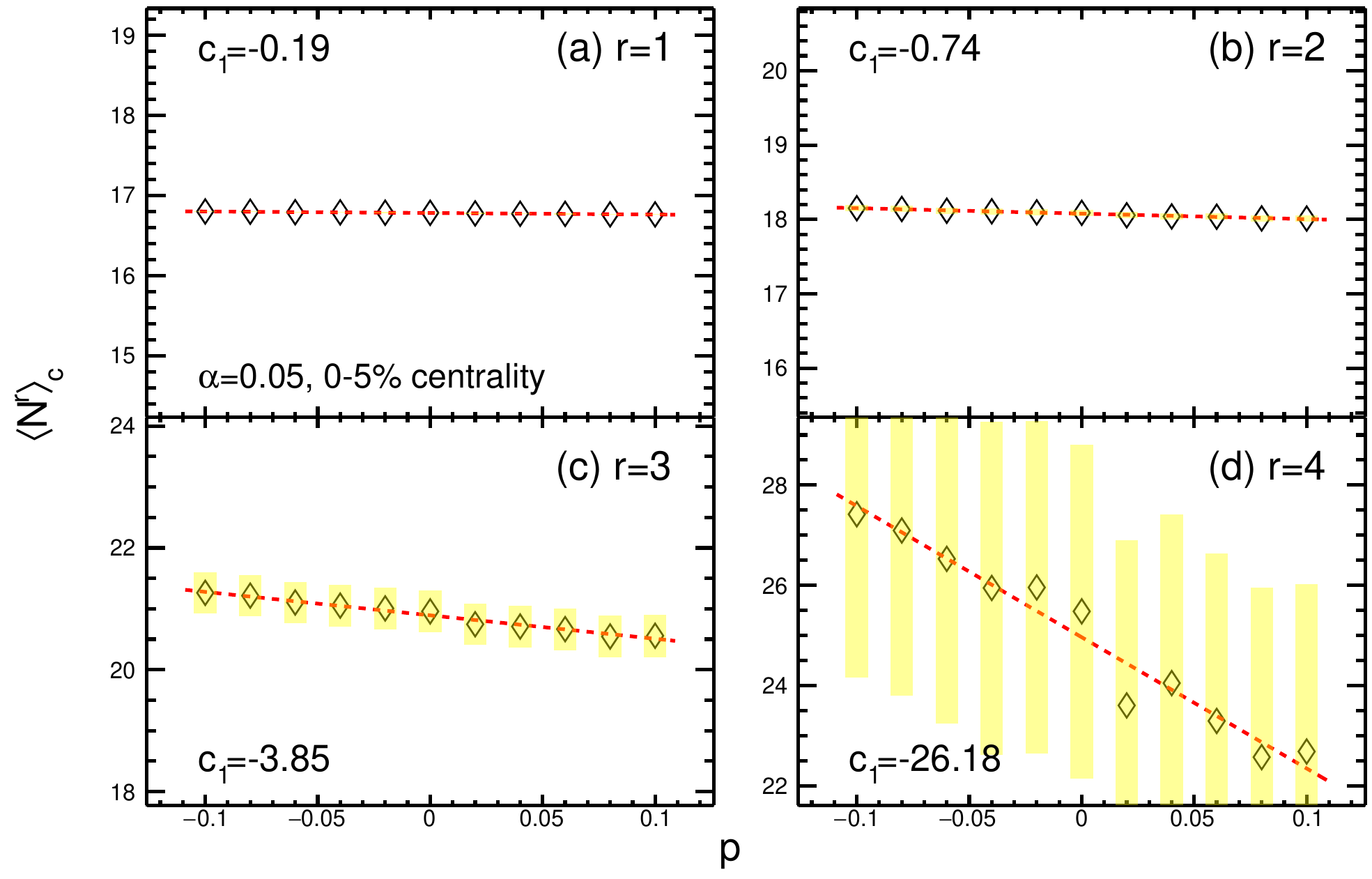}
	\end{center}
	\caption{
		Cumulants up to the $4$th-order corrected with a wrong pileup probability 
		, $\alpha(1+p)$, as a function of $p$.
		Statistical uncertainties are shown in bands. 
		The red dotted lines are the polynomial fit functions with $c_{1}p+c_{0}$. 
		The values of the fit parameter $c_{1}$ are shown in the panel.
		The scale of y-axis is set to $\pm$15\% with respect 
		to the cumulant values at $p=0$ for all panels.
		}
	\label{fig:sysalpha}
\end{figure}

To check how the uncertainty of $w_{i,j}$ and $\alpha_m$ affects the
final result, we again employ the model in Sec.~\ref{sec:realistic}
and perform the pileup correction using
wrong pileup probabilities, 
$\alpha(1+p)$, with $\alpha=0.05$.
We vary the value of $p$ from -10\% to 10\% and determine the values of $w_{i,j}$ and
$\alpha_m$ according to Eqs.~\ref{eq:w_ij} and \ref{eq:alpha_m}.
The pileup correction is then performed with these wrong probabilities.
Figure~\ref{fig:sysalpha} shows
cumulants up to the $4$th order at 0-5\% centrality
as functions of $p$.
It can be found that the results are overcorrected for $p>0$, 
while the corrections are not enough for $p<0$. 
Further, higher-order cumulants get more affected by wrong values of 
the correction parameters
as seen in the larger slope of the fitted functions.
We would propose to consider those variations as systematic uncertainties on final results.

\section{Summary\label{sec:summary}}
In this paper, we proposed a method to correct the effect
of the pileup events on the higher-order moments and cumulants.
The method can be derived by decomposing pileups into 
various combinations of sub-pileup events in terms of moments. 
The moments for sub-pileup events can be reconstructed assuming 
that the pileups are the consequences of the superposition between two independent events.
We utilized the fact that the pileup changes the total multiplicity. 
The correction formulas are expressed by the sub-pileup moments and 
the moments from the lower multiplicity events, thus solvable from the lowest 
multiplicity events. Two models are performed with the same mean values 
of particle distributions for all multiplicity events, and with the $N_{\rm part}$-dependent
mean values. The method works correctly for both cases.
The method can deal with the pileup events for 
more than two single-collisions.
The effect of trigger inefficiencies needs to be carefully checked  
by changing the starting point of the recursive corrections.
The systematic uncertainties will be reduced by determining 
the pileup probability precisely. 

Finally, we remark that one has to make sure that the detector efficiencies are 
corrected in a proper way
~\cite{Bialas:1985jb,Kitazawa:2011wh,eff_kitazawa,eff_koch,eff_psd_volker,eff_xiaofeng,eff_psd_kitazawa,binomial_breaking,Nonaka:2017kko,Kitazawa:2017ljq,Nonaka:2018mgw,Esumi:2020xdo} before performing the pileup correction. 

\section{Acknowledgement}
This work was supported by Ito Science Foundation (2017) and
JSPS KAKENHI Grant No. 25105504, 17K05442 and 19H05598.

\bibliography{main}

\end{document}